\documentclass[aps, prl, twocolumn, showpacs, ams math, assymb, superscriptaddress, floatfix]{revtex4-1}
\usepackage{hyperref}
\usepackage{graphicx}
\usepackage{float}

\begin{document}

\title{Identification of Deeply Bound Heteronuclear Molecules Using Pulsed Laser Depletion Spectroscopy}

\author{P. Zabawa}
\author{A. Wakim}
\author{A. Neukirch}
\author{C. Haimberger}
\author{N. P. Bigelow}
\affiliation{Department of Physics and Astronomy, University of Rochester, Rochester NY 14627}
\author{A. V. Stolyarov}
\author{E. A. Pazyuk}
\affiliation{Department of Chemistry, Moscow State University, 119991, Moscow, Russia}
\author{M. Tamanis}
\author{R. Ferber}
\affiliation{Laser Center, University of Latvia, 19 Rainis Blvd., Riga LV-1586, Latvia}

\begin{abstract}

We demonstrate that a near-dissociation photoassociation resonance can be used to create a deeply bound molecular sample of ultracold NaCs. To probe the resulting vibrational distribution of the sample, we use a new technique that can be applied to any ultracold molecular system. We utilize a tunable pulsed dye laser to produce efficient spectroscopic scans ($\sim700$ cm$^{-1}$ at a time) in which we observe the $1^{1} \protect\Sigma^{+}\protect\rightarrow 2^{1}\protect\Sigma^{+}-2^{3}\protect\Pi$ vibrational progression, as well as the dissociation limit to the Cs 6$^{2}$P$_{3/2}$ asymptote. We assign $1^{1} \protect\Sigma^{+}$$(\emph{v}$ = 4, 5, 6, 11, 19) vibrational levels in our sample.

\end{abstract}

\maketitle

The creation and investigation of ultracold heteronuclear molecules has attracted significant interest.  When preparing molecular samples, particular emphasis has been placed on populating deeply bound levels in the singlet ground state because those rovibrational states are highly polarizable and can be used in experiments that explore the resulting strong dipole-dipole (``polar molecule'') interactions. For example, ultracold polar molecules can be used as qubits and in quantum memories \cite{cote2006, *lee2005, *demille2002}, for quantum simulation \cite{nagerl2010}, and to study strongly interacting quantum degenerate gases \cite{ni2009}.

There have been a number of recent successes in creating ultracold absolute ground state heteronuclear molecules starting from trapped atomic vapors. In the case of RbCs \cite{demille2005}, atoms were photoassociated into excited molecules and then transferred into the rovibrational ground state using an incoherent pump-dump laser excitation scheme. In the case of KRb \cite{ni2008}, an elegant, yet complex technique was used in which weakly bound KRb molecules were formed via a Feshbach resonance after which the molecules were transferred to the rovibrational singlet ground state using a coherent two-photon (STIRAP) process enabled by a highly stabilized fs laser frequency comb. In LiCs \cite{weidemuller2008}, absolute ground state molecules were formed by photoassociation (PA) followed by spontaneous emission directly into the ground state. The simplicity of this single PA step is an attractive technique; however, the molecular production rate of a desired state is diluted by the creation of a wide range of rovibrational ground states.

\begin{figure}
\centering
\includegraphics{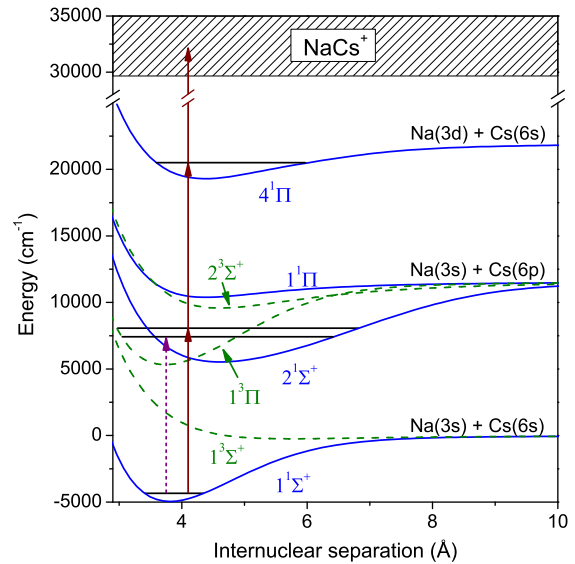}
\caption{\label{spaghetti}Energy diagram of the pulsed depletion experiment. The depletion pulse (dashed arrow) drives a resonant transition between electronic states, and the remaining sample is detected with REMPI (solid arrows). The \textit{ab initio} PECs are from \cite{korek2000}, and the ionization threshold is from\cite{korek08, *stoll82, *valance78, *ferrante74}}
\end{figure}

To help control this distribution, the PA frequency is chosen to populate an excited molecular state that preferentially decays to specific rovibrational ground states.  Once this is accomplished, purification of the sample into a single rovibrational state is achieved using any one of a variety of methods such as vibrational state cooling by optical pumping \cite{pillet2008}, collision induced vibrational relaxation \cite{hudson2008}, or state-selective electric trapping \cite{meijer2005}. What is required to optimize any of these approaches is a clear understanding of the rovibrational population distribution created by the initial PA step.

Resonance Enhanced Multi-Photon Ionization (REMPI) spectroscopy provides a powerful technique for measurement of the vibrational state distribution.  However, the single-color, two-photon REMPI spectrum for ultracold NaCs molecules is difficult to analyze due to the dense, congested nature of the spectrum and the uncertainty in potential energy curves (PEC) for the states involved.  An attractive strategy for state identification is depletion spectroscopy  \cite{wang2007, weidemuller2008}. In the cold molecule community, this is performed using a narrow linewidth $(<1$ MHz$)$ CW laser to deplete specific rovibrational states, allowing assignment of portions of the REMPI spectra. However, this technique is not suitable for mapping out large segments of the vibrational progression because the typical vibrational energy splittings ($\sim10$ to 100 cm$^{-1}$) are not well matched to the tuning characteristics of a CW depletion laser.

\begin{figure}[t!]
\centering
\includegraphics{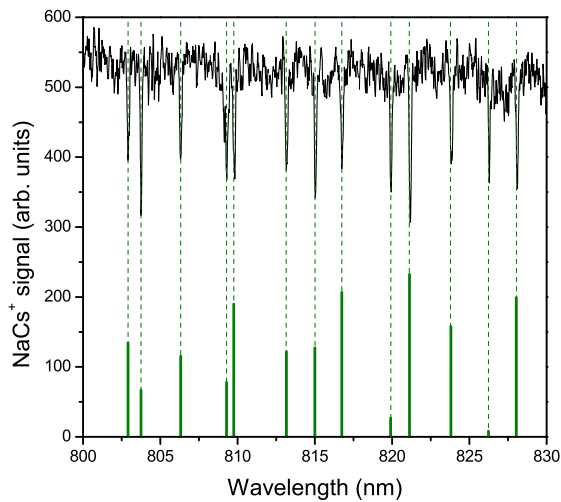}
\caption{\label{pids}Depletion scan revealing $1^{1}\Sigma^{+}$($\emph{v}$ = 4) $\rightarrow$ $2^{1}\Sigma^{+}-2^{3}\Pi$. Vertical bars are the calculated, scaled transition dipole moments.}
\end{figure}

In this paper, we introduce a pulsed-laser depletion method to identify the vibrational states in our REMPI spectrum. We thereby show that our PA process begins with a near-dissociation resonance in an $\Omega = 1$ state and leads to efficient population of deeply bound NaCs molecules in the singlet ground state.  We further compare our results to term values extrapolated from a hot-molecule Collision Enhanced Laser Induced Fluorescence (CELIF) experiment \cite{stolyarov2009}. We also observe the dissociation of $\emph{v}=5$ molecules to the Cs $6^{2}$P$_{3/2}$ asymptote yielding a direct measurement of the binding energy of the initial state, and assign spectra of the $3(\Omega = 1)$ and $4(\Omega = 1)$ electronic states.

We begin with ultracold sodium and cesium atoms $(\sim200 \, \mu$K$)$ in dark-SPOT Magneto-Optical Traps. The atoms are photoassociated with 500 mW from a Ti:Sa ring laser at a frequency that has been detuned from the Cs $6^{2}$P$_{3/2}$ atomic line; for more experimental details see \cite{chris2004}. Several PA lines detuned from this asymptote have been labeled \cite{chris2009}.

A diagram of the laser frequencies used in the experiment is given in FIG. ~\ref{spaghetti}. The REMPI detection laser is fixed to an unassigned vibrational line in the photoionization spectrum to obtain a constant baseline signal as can be seen in FIG. ~\ref{pids}. While the depletion laser is scanned at a resolution of $\sim0.6$ cm$^{-1}$, dips in the signal indicate resonances with the excited state.

\begin{figure}
\centering
\includegraphics{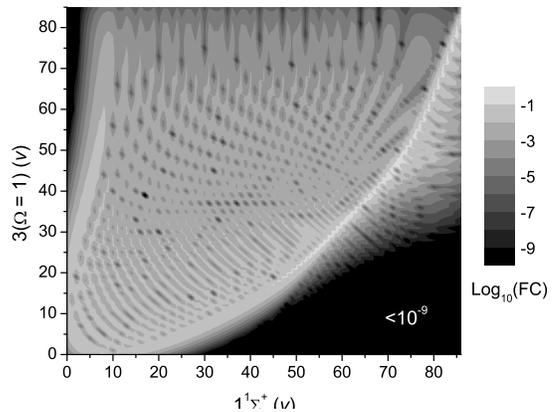}
\caption{\label{fcmaps}Franck-Condon plot of $1^{1} \Sigma^{+} \rightarrow 4(\Omega = 1)$, calculated with Level \cite{Roy2007} using an empirically fit potential \cite{Zaharova2007}. Note the overlap of near-dissociation vibrational levels with a wide range of $1^{1}\Sigma^{+}$ vibrational levels.}
\end{figure}

We choose a PA resonance detuned $\sim23$ GHz below the Cs $6^{2}$P$_{3/2}$ atomic line. This detuning exhibits high production rates of deeply bound molecules. To assign the near-dissociation PA line, we examine the Franck-Condon overlap between high lying vibrational levels in the excited electronic states and the deeply bound levels in the $1^{1} \Sigma^{+}$ state, see FIG.~\ref{fcmaps}. This electronic state is a strong candidate to produce absolute ground state molecules \cite{stwalley10}.

To calculate these Franck-Condon factors, we use the relevant \textit{ab initio} spin-orbit potentials from Korek \textit{et al.} \cite{korek2007}, and the experimentally fit $1^{1}\Pi$ state \cite{Docenko2006}. We smoothly attach long range analytic potentials \cite{Bussery1987} to the \textit{ab initio} potentials using the Fermi function \cite{napolitano2005}.

The $3(\Omega = 0^{+})$, $3(\Omega = 1)$, and $4(\Omega = 1)$ states correspond to the Cs $6^{2}$P$_{3/2}$ asymptote and have dipole allowed transitions to $1^{1} \Sigma^{+}$. However, only the $\Omega = 1$ states have good Franck-Condon overlap $(\sim10^{-5})$ for the observed transitions. Two other PA resonances previously assigned as $\Omega = 1$ \cite{chris2009} also produce deeply bound molecules. PA lines corresponding to the $1(\Omega=2)$ electronic state do not yield molecules with $\emph{v}$ $<$ $30$.

During the experiment, the PA beam is extinguished 100 $\mu$s before the 10 ns depletion laser pulse is introduced. After another 100 $\mu$s, a second pulse ionizes the remaining sample. With a pulse intensity of 6$\times 10^{6}$ W/cm$^{2}$, we achieve more than $50\%$ depletion of the ion signal on resonance, limited by spatial mode-matching of the detection and depletion beams. Both the depletion and REMPI detection lasers are independent tunable pulsed dye lasers. Ion species are resolved and detected with a time-of-flight mass spectrometer.

We isolate ground vibrational levels in the REMPI spectrum by choosing a three photon, single-color ionization pathway using the heavily mixed $2^{1}\Sigma^{+}-2^{3}\Pi$ complex as the first intermediate state. The dependence on a second resonant excitation for ionization modulates the spectrum and decreases the probability that a single frequency will ionize multiple ground states. This modulation may suppress specific vibrational levels whose second transitions are not resonant and therefore much less likely to occur.

Throughout this paper, we will refer to the ground electronic states in the Hund's case (a) notation, and the heavily mixed excited electronic states are given in the Hund's case (c) basis, or as a fully perturbed complex. The $3(\Omega = 1)$ and $4(\Omega = 1)$ states correspond roughly to the $2^{3}\Sigma^{+}$ and $1^{1}\Pi$, respectively, but we emphasize the fact that both observed states have dipole allowed transitions to the $1^{1}\Sigma^{+}$ ground state.

In FIG.~\ref{pids}, we display the remarkable agreement between our pulsed depletion spectra and extrapolations from the CELIF data. This technique allows us to observe vibrational progressions from any detectable state in the REMPI spectrum efficiently, and label them precisely. The calibration of the depletion laser was performed using well-known Cs atomic transitions \cite{KLEIMAN1962}, and we find a root mean squared deviation (RMSD) of 0.56 cm$^{-1}$ between the 45 assigned and calculated term values. We detect significant populations in the $\emph{v} =$ 4, 5, 6, 11 and 19 vibrational states. All of these vibrational levels and any with $\emph{v}$ $<$ $30$ have a permanent electric dipole moment of $\sim4.6$ Debye \cite{dulieu2005}.

\begin{figure}
\centering
\includegraphics{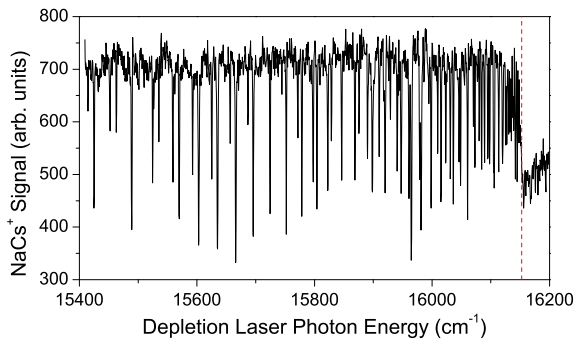}
\caption{\label{dlimit}Depletion scan revealing resonances between $1^{1}\Sigma^{+}$ ($\emph{v}=5$) and the excited vibrational levels below the Cs $6^{2}$P$_{3/2}$ asymptote. The dashed line indicates the calculated value of the dissociation energy.}
\end{figure}

\begin{figure}
\centering
\includegraphics{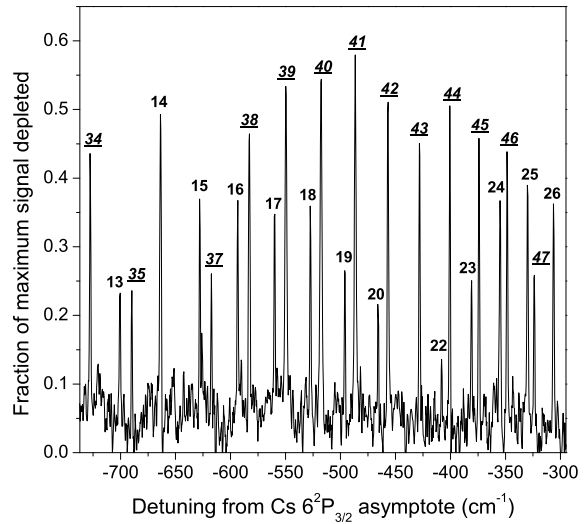}
\caption{\label{numbered}Assignments to the $3(\Omega = 1)$ (underlined) and $4(\Omega = 1)$ states. This spectrum is an inverted portion of FIG.~\ref{dlimit}.}
\end{figure}

Using the pulsed depletion technique, we detect dissociation of NaCs along the 6$^{2}$P$_{3/2}$ asymptote starting from the $1^{1}\Sigma^{+}$ (\emph{v} = 5) state, as shown in FIG.~\ref{dlimit}. This allows us to extract a binding energy without any \emph{a priori} knowledge of the molecular structure. We scan the depletion laser across the Cs $6^{2}$P$_{3/2}$ dissociation asymptote, and the last discernible vibronic transition occurs at 16150.5 cm$^{-1}$, setting a lower bound on the binding energy. To estimate an upper bound, we fit the step near $\sim$16150 cm$^{-1}$ in FIG.~\ref{dlimit} and find the cutoff at 16156.1 cm$^{-1}$.  This places the experimental range of the binding energy between 4418.2 and 4423.8 cm$^{-1}$. The predicted value for the binding energy is 4420.3 cm$^{-1}$, as calculated from an experimentally fit $1^{1}\Sigma^{+}$ PEC \cite{Zaharova2007} using LEVEL \cite{Roy2007}, and falls within our experimental upper and lower bounds.xzv 

We assign the observed vibrational spectra in FIG.~\ref{dlimit} up to $\sim$16000 cm$^{-1}$. We do not attempt to label states closer to the asymptote because perturbations from overlapping states would affect the integrity of the analysis. The resulting assignments are shown in FIG.~\ref{numbered}. To make the assignments, we compare calculated and observed term values, Franck-Condon modulation and vibrational energy spacings. For the $4(\Omega = 1)$, we use the experimental $1^{1}\Pi$ state \cite{Zaharova2007}, and calculate the predicted spectrum using LEVEL. We find a RMSD of 2.3 cm$^{-1}$ between the observed and calculated term values.  The calculated average vibrational energy spacings of 25.4 cm$^{-1}$ matches well with the observed spacings of 25.8 cm$^{-1}$. Note that some of the vibrational states described here are outside the range (\emph{v} $>$ 25) of the experimentally fit PEC \cite{Zaharova2007}.

For the $3(\Omega = 1)$ electronic state, we compare our term values with those calculated from the \emph{ab initio} potential given in \cite{korek2007}. We find a RMSD between observed and calculated term values of 9.6 cm$^{-1}$. The average vibrational energy spacings are 22.5 cm$^{-1}$ for the \emph{ab initio} PEC, compared to the observed spacings of 24.1 cm$^{-1}$.

In summary, we present a near-dissociation PA resonance which populates deeply bound states assigned as 1$^{1}\Sigma^{+}$(\emph{v} = 4, 5, 6, 11, 19). Previous single-step PA experiments utilizing weakly bound excited states did not report production of such a wide range of deeply bound molecules \cite{mancini2004, *kerman2004}. We describe a pulsed depletion technique and show that it is an efficient method for performing vibrational spectroscopy on a sample of ultracold molecules. 

In future work, we intend to probe deeper resonances ($\sim$100 cm$^{-1}$) in the 6$^{2}$P$_{3/2}$ PA manifold to directly produce \emph{v} = 0 molecules. Detection of these lines will be aided by our investigation of the $3(\Omega=1)$ and $4(\Omega=1)$ electronic states with pulsed laser depletion spectroscopy. A combination of the pulsed and CW depletion spectroscopic techniques will facilitate efficient monitoring of state populations in electrostatic or ac trapping \cite{kleinert2007,*twist2007} and vibrational state cooling experiments.

This work was supported by the NSF and ARO. The Moscow team acknowledges support from the Russian Foundation for Basic Researches (grant N10-03-00195-a) and MSU Priority Direction 2.3.

\bibliography{pulseddepletionpaper15}

\begin{thebibliography}{10}%
\makeatletter
\providecommand \@ifxundefined [1]{%
 \ifx #1\undefined \expandafter \@firstoftwo
 \else \expandafter \@secondoftwo
\fi
}%
\providecommand \@ifnum [1]{%
 \ifnum #1\expandafter \@firstoftwo
 \else \expandafter \@secondoftwo
\fi
}%
\providecommand \enquote [1]{``#1''}%
\providecommand \bibnamefont  [1]{#1}%
\providecommand \bibfnamefont [1]{#1}%
\providecommand \citenamefont [1]{#1}%
\providecommand\href[0]{\@sanitize\@href}%
\providecommand\@href[1]{\endgroup\@@startlink{#1}\endgroup\@@href}%
\providecommand\@@href[1]{#1\@@endlink}%
\providecommand \@sanitize [0]{\begingroup\catcode`\&12\catcode`\#12\relax}%
\@ifxundefined \pdfoutput {\@firstoftwo}{%
 \@ifnum{\z@=\pdfoutput}{\@firstoftwo}{\@secondoftwo}%
}{%
 \providecommand\@@startlink[1]{\leavevmode\special{html:<a href="#1">}}%
 \providecommand\@@endlink[0]{\special{html:</a>}}%
}{%
 \providecommand\@@startlink[1]{%
  \leavevmode
  \pdfstartlink
   attr{/Border[0 0 1 ]/H/I/C[0 1 1]}%
   user{/Subtype/Link/A<</Type/Action/S/URI/URI(#1)>>}%
  \relax
 }%
 \providecommand\@@endlink[0]{\pdfendlink}%
}%
\providecommand \url  [0]{\begingroup\@sanitize \@url }%
\providecommand \@url [1]{\endgroup\@href {#1}{\urlprefix}}%
\providecommand \urlprefix [0]{URL }%
\providecommand \Eprint[0]{\href }%
\@ifxundefined \urlstyle {%
  \providecommand \doi [1]{doi:\discretionary{}{}{}#1}%
}{%
  \providecommand \doi [0]{doi:\discretionary{}{}{}\begingroup
  \urlstyle{rm}\Url }%
}%
\providecommand \doibase [0]{http://dx.doi.org/}%
\providecommand \Doi[1]{\href{\doibase#1}}%
\providecommand \bibAnnote [3]{%
  \BibitemShut{#1}%
  \begin{quotation}\noindent
    \textsc{Key:}\ #2\\\textsc{Annotation:}\ #3%
  \end{quotation}%
}%
\providecommand \bibAnnoteFile [2]{%
  \IfFileExists{#2}{\bibAnnote {#1} {#2} {\input{#2}}}{}%
}%
\providecommand \typeout [0]{\immediate \write \m@ne }%
\providecommand \selectlanguage [0]{\@gobble}%
\providecommand \bibinfo [0]{\@secondoftwo}%
\providecommand \bibfield [0]{\@secondoftwo}%
\providecommand \translation [1]{[#1]}%
\providecommand \BibitemOpen[0]{}%
\providecommand \bibitemStop [0]{}%
\providecommand \bibitemNoStop [0]{.\EOS\space}%
\providecommand \EOS [0]{\spacefactor3000\relax}%
\providecommand \BibitemShut [1]{\csname bibitem#1\endcsname}%
\bibitem{cote2006}%
  \BibitemOpen
  \bibfield{author}{%
  \bibinfo {author} {\bibfnamefont{S.~F.}\ \bibnamefont{Yelin}}, \bibinfo
  {author} {\bibfnamefont{K.}~\bibnamefont{Kirby}},\ and\ \bibinfo {author}
  {\bibfnamefont{R.}~\bibnamefont{C\^ot\'e}},\ }%
  \bibfield{journal}{%
  \Doi{10.1103/PhysRevA.74.050301}{\bibinfo {journal} {Phys. Rev. A}}\ }%
  \textbf{\bibinfo {volume} {74}},\ \bibinfo {pages} {050301} (\bibinfo {year}
  {2006})%
  \bibAnnoteFile{NoStop}{cote2006}%
\bibitem{lee2005}%
  \BibitemOpen
  \bibfield{author}{%
  \bibinfo {author} {\bibfnamefont{C.}~\bibnamefont{Lee}}\ and\ \bibinfo
  {author} {\bibfnamefont{E.~A.}\ \bibnamefont{Ostrovskaya}},\ }%
  \bibfield{journal}{%
  \Doi{10.1103/PhysRevA.72.062321}{\bibinfo {journal} {Phys. Rev. A}}\ }%
  \textbf{\bibinfo {volume} {72}},\ \bibinfo {pages} {062321} (\bibinfo {year}
  {2005})%
  \bibAnnoteFile{NoStop}{lee2005}%
\bibitem{demille2002}%
  \BibitemOpen
  \bibfield{author}{%
  \bibinfo {author} {\bibfnamefont{D.}~\bibnamefont{DeMille}},\ }%
  \bibfield{journal}{%
  \Doi{10.1103/PhysRevLett.88.067901}{\bibinfo {journal} {Phys. Rev. Lett.}}\
  }%
  \textbf{\bibinfo {volume} {88}},\ \bibinfo {pages} {067901} (\bibinfo {year}
  {2002})%
  \bibAnnoteFile{NoStop}{demille2002}%
\bibitem{nagerl2010}%
  \BibitemOpen
  \bibfield{author}{%
  \bibinfo {author} {\bibfnamefont{J.~G.}\ \bibnamefont{Danzl}}, \bibinfo
  {author} {\bibfnamefont{M.~J.}\ \bibnamefont{Mark}}, \bibinfo {author}
  {\bibfnamefont{E.}~\bibnamefont{Haller}}, \bibinfo {author}
  {\bibfnamefont{M.}~\bibnamefont{Gustavsson}}, \bibinfo {author}
  {\bibfnamefont{R.}~\bibnamefont{Hart}}, \bibinfo {author}
  {\bibfnamefont{J.}~\bibnamefont{Aldegund}}, \bibinfo {author}
  {\bibfnamefont{J.~M.}\ \bibnamefont{Hutson}},\ and\ \bibinfo {author}
  {\bibfnamefont{H.-C.}\ \bibnamefont{N\"agerl}},\ }%
  \bibfield{journal}{%
  \bibinfo {journal} {Nature Physics}\ }%
  \textbf{\bibinfo {volume} {6}},\ \bibinfo {pages} {265} (\bibinfo {year}
  {2010})%
  \bibAnnoteFile{NoStop}{nagerl2010}%
\bibitem{ni2009}%
  \BibitemOpen
  \bibfield{author}{%
  \bibinfo {author} {\bibfnamefont{S.}~\bibnamefont{Ospelkaus}}, \bibinfo
  {author} {\bibfnamefont{K.-K.}\ \bibnamefont{Ni}}, \bibinfo {author}
  {\bibfnamefont{M.~H.~G.}\ \bibnamefont{de~Miranda}}, \bibinfo {author}
  {\bibfnamefont{B.}~\bibnamefont{Neyenhuis}}, \bibinfo {author}
  {\bibfnamefont{D.}~\bibnamefont{Wang}}, \bibinfo {author}
  {\bibfnamefont{S.}~\bibnamefont{Kotochigova}}, \bibinfo {author}
  {\bibfnamefont{P.}~\bibnamefont{Julienne}}, \bibinfo {author}
  {\bibfnamefont{D.~S.}\ \bibnamefont{Jin}},\ and\ \bibinfo {author}
  {\bibfnamefont{J.}~\bibnamefont{Ye}},\ }%
  \bibfield{journal}{%
  \bibinfo {journal} {Faraday Discuss.}\ }%
  \textbf{\bibinfo {volume} {142}},\ \bibinfo {pages} {351} (\bibinfo {year}
  {2009})%
  \bibAnnoteFile{NoStop}{ni2009}%
\bibitem{demille2005}%
  \BibitemOpen
  \bibfield{author}{%
  \bibinfo {author} {\bibfnamefont{J.~M.}\ \bibnamefont{Sage}}, \bibinfo
  {author} {\bibfnamefont{S.}~\bibnamefont{Sainis}}, \bibinfo {author}
  {\bibfnamefont{T.}~\bibnamefont{Bergeman}},\ and\ \bibinfo {author}
  {\bibfnamefont{D.}~\bibnamefont{DeMille}},\ }%
  \bibfield{journal}{%
  \bibinfo {journal} {Phys. Rev. Lett.}\ }%
  \textbf{\bibinfo {volume} {94}},\ \bibinfo {pages} {203001} (\bibinfo {year}
  {2005})%
  \bibAnnoteFile{NoStop}{demille2005}%
\bibitem{ni2008}%
  \BibitemOpen
  \bibfield{author}{%
  \bibinfo {author} {\bibfnamefont{K.-K.}\ \bibnamefont{Ni}}, \bibinfo {author}
  {\bibfnamefont{S.}~\bibnamefont{Ospelkaus}}, \bibinfo {author}
  {\bibfnamefont{M.~H.~G.}\ \bibnamefont{de~Miranda}}, \bibinfo {author}
  {\bibfnamefont{A.}~\bibnamefont{Pe'er}}, \bibinfo {author}
  {\bibfnamefont{B.~N. J.~J.}\ \bibnamefont{Zirbel}}, \bibinfo {author}
  {\bibfnamefont{S.}~\bibnamefont{Kotochigova}}, \bibinfo {author}
  {\bibfnamefont{P.}~\bibnamefont{Julienne}}, \bibinfo {author}
  {\bibfnamefont{D.~S.}\ \bibnamefont{Jin}},\ and\ \bibinfo {author}
  {\bibfnamefont{J.}~\bibnamefont{Ye}},\ }%
  \bibfield{journal}{%
  \bibinfo {journal} {Science}\ }%
  \textbf{\bibinfo {volume} {322}},\ \bibinfo {pages} {231} (\bibinfo {year}
  {2008})%
  \bibAnnoteFile{NoStop}{ni2008}%
\bibitem{weidemuller2008}%
  \BibitemOpen
  \bibfield{author}{%
  \bibinfo {author} {\bibfnamefont{J.}~\bibnamefont{Deiglmayr}}, \bibinfo
  {author} {\bibfnamefont{A.}~\bibnamefont{Grochola}}, \bibinfo {author}
  {\bibfnamefont{M.}~\bibnamefont{Repp}}, \bibinfo {author}
  {\bibfnamefont{K.}~\bibnamefont{M\"ortlbauer}}, \bibinfo {author}
  {\bibfnamefont{C.}~\bibnamefont{Gl\"uck}}, \bibinfo {author}
  {\bibfnamefont{J.}~\bibnamefont{Lange}}, \bibinfo {author}
  {\bibfnamefont{O.}~\bibnamefont{Dulieu}}, \bibinfo {author}
  {\bibfnamefont{R.}~\bibnamefont{Wester}},\ and\ \bibinfo {author}
  {\bibfnamefont{M.}~\bibnamefont{Weidem\"uller}},\ }%
  \bibfield{journal}{%
  \Doi{10.1103/PhysRevLett.101.133004}{\bibinfo {journal} {Phys. Rev. Lett.}}\
  }%
  \textbf{\bibinfo {volume} {101}},\ \bibinfo {pages} {133004} (\bibinfo {year}
  {2008})%
  \bibAnnoteFile{NoStop}{weidemuller2008}%
\bibitem{korek2000}%
  \BibitemOpen
  \bibfield{author}{%
  \bibinfo {author} {\bibfnamefont{M.}~\bibnamefont{Korek}}, \bibinfo {author}
  {\bibfnamefont{A.~R.}\ \bibnamefont{Allouche}}, \bibinfo {author}
  {\bibfnamefont{K.}~\bibnamefont{Fakhreddine}},\ and\ \bibinfo {author}
  {\bibfnamefont{A.}~\bibnamefont{Chaalan}},\ }%
  \bibfield{journal}{%
  \bibinfo {journal} {Can. J. Phys.}\ }%
  \textbf{\bibinfo {volume} {78}},\ \bibinfo {pages} {977} (\bibinfo {year}
  {2000})%
  \bibAnnoteFile{NoStop}{korek2000}%
\bibitem{korek08}%
  \BibitemOpen
  \bibfield{author}{%
  \bibinfo {author} {\bibfnamefont{M.}~\bibnamefont{Korek}}, \bibinfo {author}
  {\bibfnamefont{K.}~\bibnamefont{Badreddine}},\ and\ \bibinfo {author}
  {\bibfnamefont{A.~R.}\ \bibnamefont{Allouche}},\ }%
  \bibfield{journal}{%
  \bibinfo {journal} {Can. J. Phys.}\ }%
  \textbf{\bibinfo {volume} {86}},\ \bibinfo {pages} {1015} (\bibinfo {year}
  {2008})%
  \bibAnnoteFile{NoStop}{korek08}%
\bibitem{stoll82}%
  \BibitemOpen
  \bibfield{author}{%
  \bibinfo {author} {\bibfnamefont{L.}~\bibnamefont{von Szentpaly}}, \bibinfo
  {author} {\bibfnamefont{P.}~\bibnamefont{Fuentealba}}, \bibinfo {author}
  {\bibfnamefont{H.}~\bibnamefont{Preuss}},\ and\ \bibinfo {author}
  {\bibfnamefont{H.}~\bibnamefont{Stoll}},\ }%
  \bibfield{journal}{%
  \bibinfo {journal} {Chem. Phys. Lett.}\ }%
  \textbf{\bibinfo {volume} {93}},\ \bibinfo {pages} {555} (\bibinfo {year}
  {1982})%
  \bibAnnoteFile{NoStop}{stoll82}%
\bibitem{valance78}%
  \BibitemOpen
  \bibfield{author}{%
  \bibinfo {author} {\bibfnamefont{A.}~\bibnamefont{Valance}},\ }%
  \bibfield{journal}{%
  \bibinfo {journal} {J. Chem. Phys.}\ }%
  \textbf{\bibinfo {volume} {69}},\ \bibinfo {pages} {355} (\bibinfo {year}
  {1978})%
  \bibAnnoteFile{NoStop}{valance78}%
\bibitem{ferrante74}%
  \BibitemOpen
  \bibfield{author}{%
  \bibinfo {author} {\bibfnamefont{L.}~\bibnamefont{Bellomonte}}, \bibinfo
  {author} {\bibfnamefont{P.}~\bibnamefont{Cavaliere}},\ and\ \bibinfo {author}
  {\bibfnamefont{G.}~\bibnamefont{Ferrante}},\ }%
  \bibfield{journal}{%
  \bibinfo {journal} {J. Chem. Phys.}\ }%
  \textbf{\bibinfo {volume} {61}},\ \bibinfo {pages} {3225} (\bibinfo {year}
  {1974})%
  \bibAnnoteFile{NoStop}{ferrante74}%
\bibitem{pillet2008}%
  \BibitemOpen
  \bibfield{author}{%
  \bibinfo {author} {\bibfnamefont{M.}~\bibnamefont{Viteau}}, \bibinfo {author}
  {\bibfnamefont{A.}~\bibnamefont{Chotia}}, \bibinfo {author}
  {\bibfnamefont{M.}~\bibnamefont{Allegrini}}, \bibinfo {author}
  {\bibfnamefont{N.}~\bibnamefont{Bouloufa}}, \bibinfo {author}
  {\bibfnamefont{O.}~\bibnamefont{Dulieu}}, \bibinfo {author}
  {\bibfnamefont{D.}~\bibnamefont{Comparat}},\ and\ \bibinfo {author}
  {\bibfnamefont{P.}~\bibnamefont{Pillet}},\ }%
  \bibfield{journal}{%
  \Doi{10.1126/science.1159496}{\bibinfo {journal} {Science}}\ }%
  \textbf{\bibinfo {volume} {321}},\ \bibinfo {pages} {232} (\bibinfo {year}
  {2008})%
  \bibAnnoteFile{NoStop}{pillet2008}%
\bibitem{hudson2008}%
  \BibitemOpen
  \bibfield{author}{%
  \bibinfo {author} {\bibfnamefont{E.~R.}\ \bibnamefont{Hudson}}, \bibinfo
  {author} {\bibfnamefont{N.~B.}\ \bibnamefont{Gilfoy}}, \bibinfo {author}
  {\bibfnamefont{S.}~\bibnamefont{Kotochigova}}, \bibinfo {author}
  {\bibfnamefont{J.~M.}\ \bibnamefont{Sage}},\ and\ \bibinfo {author}
  {\bibfnamefont{D.}~\bibnamefont{DeMille}},\ }%
  \bibfield{journal}{%
  \bibinfo {journal} {Phys. Rev. Lett.}\ }%
  \textbf{\bibinfo {volume} {100}},\ \bibinfo {pages} {203201} (\bibinfo {year}
  {2008})%
  \bibAnnoteFile{NoStop}{hudson2008}%
\bibitem{meijer2005}%
  \BibitemOpen
  \bibfield{author}{%
  \bibinfo {author} {\bibfnamefont{J.}~\bibnamefont{van Veldhoven}}, \bibinfo
  {author} {\bibfnamefont{H.~L.}\ \bibnamefont{Bethlem}},\ and\ \bibinfo
  {author} {\bibfnamefont{G.}~\bibnamefont{Meijer}},\ }%
  \bibfield{journal}{%
  \bibinfo {journal} {Phys. Rev. Lett.}\ }%
  \textbf{\bibinfo {volume} {94}},\ \bibinfo {pages} {083001} (\bibinfo {year}
  {2005})%
  \bibAnnoteFile{NoStop}{meijer2005}%
\bibitem{wang2007}%
  \BibitemOpen
  \bibfield{author}{%
  \bibinfo {author} {\bibfnamefont{D.}~\bibnamefont{Wang}}, \bibinfo {author}
  {\bibfnamefont{J.~T.}\ \bibnamefont{Kim}}, \bibinfo {author}
  {\bibfnamefont{C.}~\bibnamefont{Ashbaugh}}, \bibinfo {author}
  {\bibfnamefont{E.~E.}\ \bibnamefont{Eyler}}, \bibinfo {author}
  {\bibfnamefont{P.~L.}\ \bibnamefont{Gould}},\ and\ \bibinfo {author}
  {\bibfnamefont{W.~C.}\ \bibnamefont{Stwalley}},\ }%
  \bibfield{journal}{%
  \Doi{10.1103/PhysRevA.75.032511}{\bibinfo {journal} {Phys. Rev. A}}\ }%
  \textbf{\bibinfo {volume} {75}},\ \bibinfo {pages} {032511} (\bibinfo {year}
  {2007})%
  \bibAnnoteFile{NoStop}{wang2007}%
\bibitem{stolyarov2009}%
  \BibitemOpen
  \bibfield{author}{%
  \bibinfo {author} {\bibfnamefont{J.}~\bibnamefont{Zaharova}}, \bibinfo
  {author} {\bibfnamefont{M.}~\bibnamefont{Tamanis}}, \bibinfo {author}
  {\bibfnamefont{R.}~\bibnamefont{Ferber}}, \bibinfo {author}
  {\bibfnamefont{A.~N.}\ \bibnamefont{Drozdova}}, \bibinfo {author}
  {\bibfnamefont{E.~A.}\ \bibnamefont{Pazyuk}},\ and\ \bibinfo {author}
  {\bibfnamefont{A.~V.}\ \bibnamefont{Stolyarov}},\ }%
  \bibfield{journal}{%
  \Doi{10.1103/PhysRevA.79.012508}{\bibinfo {journal} {Phys. Rev. A}}\ }%
  \textbf{\bibinfo {volume} {79}},\ \bibinfo {pages} {012508} (\bibinfo {year}
  {2009})%
  \bibAnnoteFile{NoStop}{stolyarov2009}%
\bibitem{chris2004}%
  \BibitemOpen
  \bibfield{author}{%
  \bibinfo {author} {\bibfnamefont{C.}~\bibnamefont{Haimberger}}, \bibinfo
  {author} {\bibfnamefont{J.}~\bibnamefont{Kleinert}}, \bibinfo {author}
  {\bibfnamefont{M.}~\bibnamefont{Bhattacharya}},\ and\ \bibinfo {author}
  {\bibfnamefont{N.~P.}\ \bibnamefont{Bigelow}},\ }%
  \bibfield{journal}{%
  \Doi{10.1103/PhysRevA.70.021402}{\bibinfo {journal} {Phys. Rev. A}}\ }%
  \textbf{\bibinfo {volume} {70}},\ \bibinfo {pages} {021402} (\bibinfo {year}
  {2004})%
  \bibAnnoteFile{NoStop}{chris2004}%
\bibitem{chris2009}%
  \BibitemOpen
  \bibfield{author}{%
  \bibinfo {author} {\bibfnamefont{C.}~\bibnamefont{Haimberger}}, \bibinfo
  {author} {\bibfnamefont{J.}~\bibnamefont{Kleinert}}, \bibinfo {author}
  {\bibfnamefont{P.}~\bibnamefont{Zabawa}}, \bibinfo {author}
  {\bibfnamefont{A.}~\bibnamefont{Wakim}},\ and\ \bibinfo {author}
  {\bibfnamefont{N.~P.}\ \bibnamefont{Bigelow}},\ }%
  \bibfield{journal}{%
  \bibinfo {journal} {New J. Phys.}\ }%
  \textbf{\bibinfo {volume} {11}},\ \bibinfo {pages} {055042} (\bibinfo {year}
  {2009})%
  \bibAnnoteFile{NoStop}{chris2009}%
\bibitem{Roy2007}%
  \BibitemOpen
  \bibfield{author}{%
  \bibinfo {author} {\bibfnamefont{R.~J.}\ \bibnamefont{Le~Roy}},\ }%
  \enquote{\bibinfo {title} {Level 8.0: \emph{A Computer Program for Solving
  the Radial Schr\"odinger Equation for Bound and Quasibound Levels}},}\
  (\bibinfo {year} {2007}),\ \url{http://leroy.uwaterloo.ca/programps/}%
  \bibAnnoteFile{NoStop}{Roy2007}%
\bibitem{Zaharova2007}%
  \BibitemOpen
  \bibfield{author}{%
  \bibinfo {author} {\bibfnamefont{J.}~\bibnamefont{Zaharova}}, \bibinfo
  {author} {\bibfnamefont{O.}~\bibnamefont{Docenko}}, \bibinfo {author}
  {\bibfnamefont{M.}~\bibnamefont{Tamanis}}, \bibinfo {author}
  {\bibfnamefont{R.}~\bibnamefont{Ferber}}, \bibinfo {author}
  {\bibfnamefont{A.}~\bibnamefont{Pashov}}, \bibinfo {author}
  {\bibfnamefont{H.}~\bibnamefont{Knockel}},\ and\ \bibinfo {author}
  {\bibfnamefont{E.}~\bibnamefont{Tiemann}},\ }%
  \bibfield{journal}{%
  \bibinfo {journal} {J. Chem. Phys.}\ }%
  \textbf{\bibinfo {volume} {127}},\ \bibinfo {eid} {224302} (\bibinfo {year}
  {2007})%
  \bibAnnoteFile{NoStop}{Zaharova2007}%
\bibitem{stwalley10}%
  \BibitemOpen
  \bibfield{author}{%
  \bibinfo {author} {\bibfnamefont{W.~C.}\ \bibnamefont{Stwalley}}, \bibinfo
  {author} {\bibfnamefont{J.}~\bibnamefont{Banerjee}}, \bibinfo {author}
  {\bibfnamefont{M.}~\bibnamefont{Bellos}}, \bibinfo {author}
  {\bibfnamefont{R.}~\bibnamefont{Carollo}}, \bibinfo {author}
  {\bibfnamefont{M.}~\bibnamefont{Recore}},\ and\ \bibinfo {author}
  {\bibfnamefont{M.}~\bibnamefont{Mastroianni}},\ }%
  \bibfield{journal}{%
  \bibinfo {journal} {J. Chem. Phys.}\ }%
  \textbf{\bibinfo {volume} {114}},\ \bibinfo {pages} {81} (\bibinfo {year}
  {2010})%
  \bibAnnoteFile{NoStop}{stwalley10}%
\bibitem{korek2007}%
  \BibitemOpen
  \bibfield{author}{%
  \bibinfo {author} {\bibfnamefont{M.}~\bibnamefont{Korek}}, \bibinfo {author}
  {\bibfnamefont{S.}~\bibnamefont{Bleik}},\ and\ \bibinfo {author}
  {\bibfnamefont{A.~R.}\ \bibnamefont{Allouche}},\ }%
  \bibfield{journal}{%
  \bibinfo {journal} {J. Chem. Phys.}\ }%
  \textbf{\bibinfo {volume} {126}},\ \bibinfo {pages} {124313} (\bibinfo {year}
  {2007})%
  \bibAnnoteFile{NoStop}{korek2007}%
\bibitem{Docenko2006}%
  \BibitemOpen
  \bibfield{author}{%
  \bibinfo {author} {\bibfnamefont{O.}~\bibnamefont{Docenko}}, \bibinfo
  {author} {\bibfnamefont{M.}~\bibnamefont{Tamanis}}, \bibinfo {author}
  {\bibfnamefont{J.}~\bibnamefont{Zaharova}}, \bibinfo {author}
  {\bibfnamefont{R.}~\bibnamefont{Ferber}}, \bibinfo {author}
  {\bibfnamefont{A.}~\bibnamefont{Pashov}}, \bibinfo {author}
  {\bibfnamefont{H.}~\bibnamefont{Kn\"ockel}},\ and\ \bibinfo {author}
  {\bibfnamefont{E.}~\bibnamefont{Tiemann}},\ }%
  \bibfield{journal}{%
  \bibinfo {journal} {J. Phys. B}\ }%
  \textbf{\bibinfo {volume} {39}},\ \bibinfo {pages} {S929} (\bibinfo {year}
  {2006})%
  \bibAnnoteFile{NoStop}{Docenko2006}%
\bibitem{Bussery1987}%
  \BibitemOpen
  \bibfield{author}{%
  \bibinfo {author} {\bibfnamefont{B.}~\bibnamefont{Bussery}}, \bibinfo
  {author} {\bibfnamefont{Y.}~\bibnamefont{Achkar}},\ and\ \bibinfo {author}
  {\bibfnamefont{M.}~\bibnamefont{Aubert-Fr\'econ}},\ }%
  \bibfield{journal}{%
  \bibinfo {journal} {Chemical Physics}\ }%
  \textbf{\bibinfo {volume} {116}},\ \bibinfo {pages} {319 } (\bibinfo {year}
  {1987})%
  \bibAnnoteFile{NoStop}{Bussery1987}%
\bibitem{napolitano2005}%
  \BibitemOpen
  \bibfield{author}{%
  \bibinfo {author} {\bibfnamefont{A.~L.~M.}\ \bibnamefont{Zanelatto}},
  \bibinfo {author} {\bibfnamefont{E.~M.~S.}\ \bibnamefont{Ribeiro}},\ and\
  \bibinfo {author} {\bibfnamefont{R.}~\bibnamefont{d.~J.~Napolitano}},\ }%
  \bibfield{journal}{%
  \bibinfo {journal} {J. Chem. Phys.}\ }%
  \textbf{\bibinfo {volume} {123}},\ \bibinfo {pages} {014311} (\bibinfo {year}
  {2005})%
  \bibAnnoteFile{NoStop}{napolitano2005}%
\bibitem{KLEIMAN1962}%
  \BibitemOpen
  \bibfield{author}{%
  \bibinfo {author} {\bibfnamefont{H.}~\bibnamefont{Kleiman}},\ }%
  \bibfield{journal}{%
  \bibinfo {journal} {J. Opt. Soc. Am.}\ }%
  \textbf{\bibinfo {volume} {52}},\ \bibinfo {pages} {441} (\bibinfo {year}
  {1962})%
  \bibAnnoteFile{NoStop}{KLEIMAN1962}%
\bibitem{dulieu2005}%
  \BibitemOpen
  \bibfield{author}{%
  \bibinfo {author} {\bibfnamefont{M.}~\bibnamefont{Aymar}}\ and\ \bibinfo
  {author} {\bibfnamefont{O.}~\bibnamefont{Dulieu}},\ }%
  \bibfield{journal}{%
  \bibinfo {journal} {J. Chem. Phys.}\ }%
  \textbf{\bibinfo {volume} {122}},\ \bibinfo {pages} {204302} (\bibinfo {year}
  {2005})%
  \bibAnnoteFile{NoStop}{dulieu2005}%
\bibitem{mancini2004}%
  \BibitemOpen
  \bibfield{author}{%
  \bibinfo {author} {\bibfnamefont{M.~W.}\ \bibnamefont{Mancini}}, \bibinfo
  {author} {\bibfnamefont{G.~D.}\ \bibnamefont{Telles}}, \bibinfo {author}
  {\bibfnamefont{A.~R.~L.}\ \bibnamefont{Caires}}, \bibinfo {author}
  {\bibfnamefont{V.~S.}\ \bibnamefont{Bagnato}},\ and\ \bibinfo {author}
  {\bibfnamefont{L.~G.}\ \bibnamefont{Marcassa}},\ }%
  \bibfield{journal}{%
  \Doi{10.1103/PhysRevLett.92.133203}{\bibinfo {journal} {Phys. Rev. Lett.}}\
  }%
  \textbf{\bibinfo {volume} {92}},\ \bibinfo {pages} {133203} (\bibinfo {year}
  {2004})%
  \bibAnnoteFile{NoStop}{mancini2004}%
\bibitem{kerman2004}%
  \BibitemOpen
  \bibfield{author}{%
  \bibinfo {author} {\bibfnamefont{A.~J.}\ \bibnamefont{Kerman}}, \bibinfo
  {author} {\bibfnamefont{J.~M.}\ \bibnamefont{Sage}}, \bibinfo {author}
  {\bibfnamefont{S.}~\bibnamefont{Sainis}}, \bibinfo {author}
  {\bibfnamefont{T.}~\bibnamefont{Bergeman}},\ and\ \bibinfo {author}
  {\bibfnamefont{D.}~\bibnamefont{DeMille}},\ }%
  \bibfield{journal}{%
  \bibinfo {journal} {Phys. Rev. Lett.}\ }%
  \textbf{\bibinfo {volume} {92}},\ \bibinfo {pages} {033004} (\bibinfo {year}
  {2004})%
  \bibAnnoteFile{NoStop}{kerman2004}%
\bibitem{kleinert2007}%
  \BibitemOpen
  \bibfield{author}{%
  \bibinfo {author} {\bibfnamefont{J.}~\bibnamefont{Kleinert}}, \bibinfo
  {author} {\bibfnamefont{C.}~\bibnamefont{Haimberger}}, \bibinfo {author}
  {\bibfnamefont{P.~J.}\ \bibnamefont{Zabawa}},\ and\ \bibinfo {author}
  {\bibfnamefont{N.~P.}\ \bibnamefont{Bigelow}},\ }%
  \bibfield{journal}{%
  \Doi{10.1103/PhysRevLett.99.143002}{\bibinfo {journal} {Phys. Rev. Lett.}}\
  }%
  \textbf{\bibinfo {volume} {99}},\ \bibinfo {pages} {143002} (\bibinfo {year}
  {2007})%
  \bibAnnoteFile{NoStop}{kleinert2007}%
\bibitem{twist2007}%
  \BibitemOpen
  \bibfield{author}{%
  \bibinfo {author} {\bibfnamefont{J.}~\bibnamefont{Kleinert}}, \bibinfo
  {author} {\bibfnamefont{C.}~\bibnamefont{Haimberger}}, \bibinfo {author}
  {\bibfnamefont{P.~J.}\ \bibnamefont{Zabawa}},\ and\ \bibinfo {author}
  {\bibfnamefont{N.~P.}\ \bibnamefont{Bigelow}},\ }%
  \bibfield{journal}{%
  \bibinfo {journal} {Rev. Sci. Instrum.}\ }%
  \textbf{\bibinfo {volume} {78}},\ \bibinfo {pages} {113108} (\bibinfo {year}
  {2007})%
  \bibAnnoteFile{NoStop}{twist2007}%
\end{thebibliography}%

\end{document}